\documentclass[twocolumn,showpacs,amsmath,amssymb,prl]{revtex4}
\usepackage{dcolumn}
\usepackage{epsfig}
\usepackage[latin1]{inputenc} 
\usepackage{bm}

\begin{document}

%
%
%
\title{Angular rigidity in tetrahedral network glasses} 
\author{M. Bauchy$^1$, M. Micoulaut$^1$, M. Celino$^2$ and C. Massobrio$^3$}
\affiliation{$^1$ Laboratoire de Physique Th\'eorique de la Mati\`ere Condens\'e,
Universit\'e Pierre et Marie Curie,  Boite 121, 4, Place Jussieu, 75252
Paris Cedex 05, France\\
$^2$ENEA, Ente per le Nuove Tecnologie, l'Energia e l'Ambiente, Unite Materiali e Nuove Tecnologie,
C.R. Casaccia, CP 2400, 00100 Roma, Italy\\
$^3$Institut de Physique et de Chimie des Mat\'eriaux de Strasbourg, 23 rue du Loess, BP43, F-67034 Strasbourg Cedex 2, France}

\date{\today}
\begin{abstract}
A set of oxide and chalcogenide tetrahedral glasses are investigated using
molecular dynamics simulations.
It is shown that unlike stoichiometric selenides such as GeSe$_2$ and SiSe$_2$,
germania and silica display large standard deviations in the associated
bond angle distributions. Within bond-bending constraints theory,
this pattern can be interpreted as a manifestation of {\it {broken}} (i.e. ineffective) oxygen bond-bending constraints.
The same analysis reveals that the changes in the Ge composition affects
mostly  bending around germanium in  binary Ge-Se systems,
leaving  Se-centred bending almost unchanged. In contrast, the corresponding Se twisting (quantified by
the dihedral angle) depends on the Ge composition and is reduced when the system becomes rigid.
Our results establishes the atomic-scale foundations of the phenomelogical rigidity theory, thereby profoundly
extending its significance and impact on the structural description of network glasses.
\end{abstract}
\pacs{61.43.Fs}
\maketitle
The large variety of physico-chemical behaviours  inherent in tetrahedral network glasses
(in particular, those involving Group IV (A=Si, Ge) oxides (X=O) or chalcogenides (X=S,Se,Te))
is deeply related to the underlying network topology, i.e. the nature of the connections  
(edge- or corner-sharing) among the basic tetrahedral structural units \cite{Salmon2009,Salmon1994}.
In the search of an unifying approach, it is tempting to follow rigidity theory, that describes 
the interplay between network properties and connectivity by considering
covalent networks in very much the same fashion as mechanical trusses \cite{rigidity_book}.
This is achieved via enumeration of mechanical rigid constraints n$_c$ arising from relevant atomic
interactions, $r/2$ bond-stretching (radial) and ($2r-3$) bond-bending (angular) for a $r$-folded atom.
As a consequence, a rigidity transition at the network mean coordination number $\bar r$=2.4
is identified \cite{rigidity_book}, separating underconstrained networks
having low-frequency (floppy) deformation modes (n$_c<3$) from overconstrained ones (n$_c>3$).
\par
Accordingly, all stoichiometric compounds AX$_2$ should
be stressed rigid i.e. they have more constraints than degrees of freedom ($n_c$=3.67), with
a mean coordination number of $\bar r$=2.67 larger than the critical
$\bar r_c$=2.4 for which optimal glass formation with small enthalpic changes at the glass transition, 
is supposed to occur \cite{Phillips1979}.
\par
The legitimacy of such a picture is challenged by the observation that
silica (SiO$_2$) and germania (GeO$_2$) are found to form rather easily
glasses \cite{easily}, in contrast with the corresponding chalcogenides (e.g. GeSe$_2$) found
at the very limit of the binary glass-forming region (in e.g. Ge$_x$Se$_{1-x}$ \cite{Bool2001}).
Furthermore, oxides have a low frequency (floppy) contribution
in the vibrational density of states, suggesting that these systems are flexible \cite{Dove} or,
at least, nearly optimally constrained \cite{Wyart} (isostatic, n$_c$=3).
The latter result can be recovered in rigidity theory under the heuristic assumption
that enhanced oxygen bond-angle values  lead to {\it {broken}} angular constraints \cite{Science} reducing
n$_c$ from 3.67 to 3.0. However, a microscopic scale rationale for this specific network
behavior is lacking.

These pieces of evidence call for the basic assertions of rigidity theory, when applied to
the above oxide and chalcogenide networks, to be revisited. Here we rely  
on an atomic-scale approach (as molecular dynamics) able to substantiate and enrich the general trends
of rigidity theory via the explicit account of the details of chemical bonding. 

\par
In this Letter we bridge the gap between constraint counting algorithms and
the statistical mechanical behavior of relevant atomic-scale quantities
(angular distributions) by considering  a set of
chalcogenides and oxide glasses. 
Two issues commonly tackled within rigidity theory are  given new atomic-scale foundations. 
First, by focussing on the stoichiometric AX$_2$ composition (i.e. SiO$_2$, GeO$_2$, SiSe$_2$ and GeSe$_2$)
we show that oxide systems contain tetrahedra which act
as rigid units having a much smaller minimal angular excursion
(quantified by $\sigma_\theta$, the standard deviation of the partial bond angle
distributions) than their chalcogenide counterparts.  
By analysing in a similar fashion the binary Ge$_x$Se$_{1-x}$ system with changing compositions,
we find that changes in rigidity (increasing Ge content $x$) are noticeable in  the angular environment of germanium,
leaving the Se centred angular excursions nearly unchanged when moving from a flexible to a stressed rigid phase.
It appears that the amplitude of the angular distribution  around A atoms is 
increased in the stressed rigid phase with an increased distorsion (i.e. higher $\sigma_\theta$s) of the tetrahedra.
This result allows reconsidering the general accepted picture of a
Ge-Se network made of flexible Se-chains and rigid GeSe$_{4/2}$ tetrahedra \cite{rigidity_book}. 
As a second main outcome, we find that $\sigma_\theta$ for oxygen in oxides,
is much larger than $\sigma_\theta$ for selenium in GeSe$_2$, proving 
that oxygen bending constraints are {\it {broken}} (i.e. ineffective) in SiO$_2$ and GeO$_2$.
This provides a microscopic rationale for the sensitivity to
bond-bending around  oxygen in oxide networks, consistent with experimental
evidence \cite{Science}. 

\par\begin{figure}
\begin{center}
\epsfig{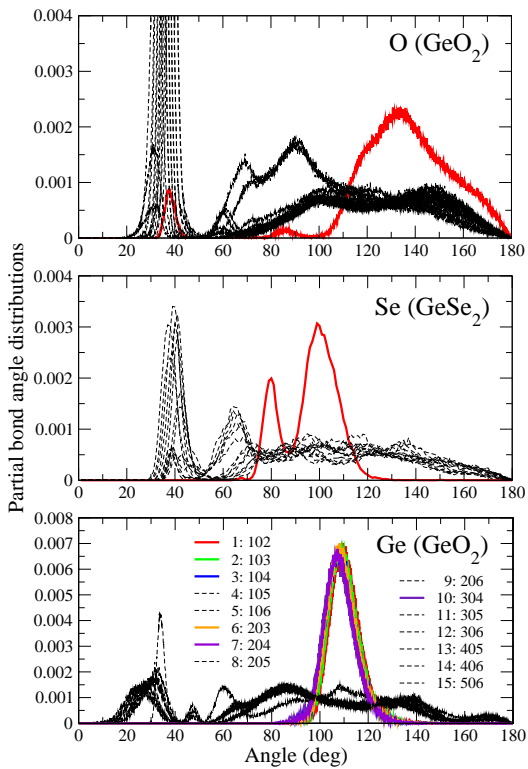}
\end{center}
\caption{(color online) From top to bottom: oxygen, selenium and germanium partial
bond angle distributions (PBAD) in GeO$_2$ and GeSe$_2$ for an arbitrary N=6.
The colored curves correspond to PBADs having the lowest standard deviation(s) $\sigma_\theta$ 
(see Fig. 2). The sharp peaks at $\theta\simeq$ 40$^o$ correspond to the hard-core repulsion.
Labels defined in the bottom panel are used throughout the text.}
\end{figure}
Our analysis is based on molecular dynamics trajectories obtained at T=300 K for
a set of glassy systems encompassing GeO$_2$, SiO$_2$, SiSe$_2$ and Ge$_x$Se$_{1-x}$ for 4 different compositions:
$x=0.10$ (GeSe$_9$ in the flexible phase), $x=0.20$ (GeSe$_4$ at the rigidity transition \cite{Phillips1979}),
$x=0.25$ (GeSe$_3$ in the intermediate phase \cite{Bool2001}) and $x=0.33$ (GeSe$_2$ in the stressed rigid phase).
Oxides at the experimental densities have been simulated  according to
Ref. \cite{PRE2006} and \cite{JCP1996} using a classical Born-Mayer force field. Due to the
large difference of electronegativity between Si(Ge) and O, this choice ensures
plausible qualitative modelling within classical molecular dynamics. In the case of SiSe$_2$ and Ge$_x$Se$_{1-x}$
systems, we resort to first-principle molecular dynamics (FPMD) within a fully self-consistent
framework that proved adequate to describe chemical bonding and its changes with concentration.
For some of the systems, temporal trajectories recorded previously  are substantially extended to
attain optimal statistical accuracy (within a few percent at most). Otherwise, as in the
case of GeSe$_9$ and GeSe$_3$,  glassy structures are produced from the outset
after cooling from the liquid state and appropriate structural relaxation.
Overall, typical time trajectories for collection of the averages in the glassy state cover $\sim$ 100  ps.     
Details on the FPMD methodology and the productions of glassy structure are
extensively reported in \cite{gese2} and \cite{sise2}. For all glassy structures, the
atomic-scale picture is consistent with that obtained by using an alternative first-principles
approach \cite{Drabold}.
\begin{figure}
\begin{center}
\epsfig{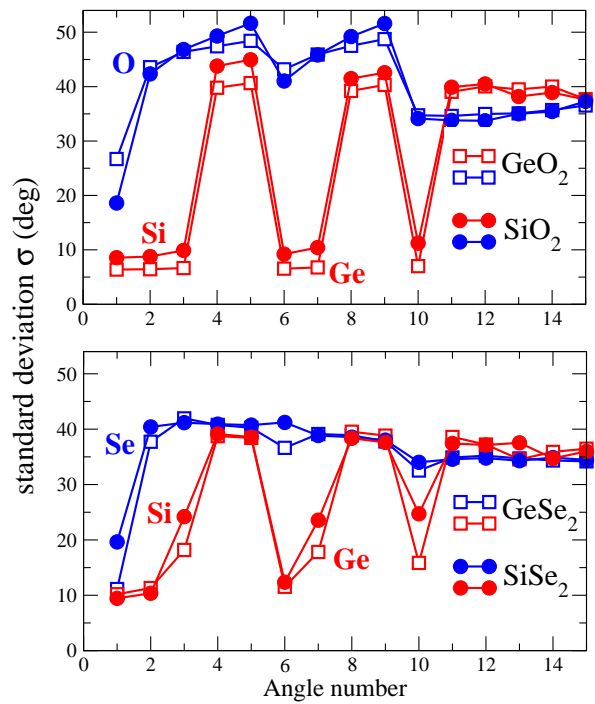}
\end{center}
\vspace{3 cm}
\caption{(color online) Standard deviation $\sigma_{\theta_{ij}}$ of
partial bond angle distributions (PBADs) in oxides (SiO$_2$, GeO$_2$, top panel)
and chalcogenides (SiSe$_2$, GeSe$_2$, bottom panel) as a
function of the angle number (see defintion in Fig. 1).}
\end{figure}
\par
To obtain  the number of bond-stretching interactions we have focused on neighbour distribution functions (NDFs). 
A set of NDFs can be defined by fixing the neighbor number $n$ (first, second etc), 
the sum of all NDFs yielding the standard pair distribution function g$_i$(r)
($i$=A,X). Integration of g$_i$(r)($i$=A,X) up to the first minimum gives the coordination numbers
r$_X$ and r$_A$, and hence the corresponding number of bond-stretching constraints r$_i$/2 ($i$=A,X).
For all considered systems, we find $r_X=2$ and $r_A=4$ leading to 1 and 2 bond-stretching constraints for the X and A 
atoms. For instance, when integrating the pair distribution function up to its first minimum, we find for GeO$_2$ $r_{Ge}$=4.01 and 
$r_{O}$=1.97, and for GeSe$_2$, $r_{Ge}$=4.02 and $r_{Se}$=1.96, in agreement with experiments \cite{Salmon1994,Gladden,Salmon}.
\par
Bond-bending constraint counting is based on partial bond angle distributions (PBADs) $P({\theta_{ij}})$ defined as follows:
for each type of central atom $0$, the N first neighbours $i$ are selected and the
N(N-1)/2 corresponding angles $i0j$ ($i$=1..N-1, $j$=2..N) computed, i.e. 102, 103, 203, etc.
The standard deviation $\sigma_{\theta_{ij}}$ of the distribution P($\theta_{ij}$) gives
a quantitative estimate about the angular excursion around a mean angular value,
and provides a measure of the bond-bending strength
(small values, i.e. intact bond-bending constraint) or the bond-bending weakness (large values,
ineffective or {\it {broken}} constraint).
\par
Figure 1 shows such PBADs for oxygen (Ge-O-Ge angle) and germanium (O-Ge-O) in GeO$_2$ and
for selenium (Ge-Se-Ge) in GeSe$_2$ for an arbitrary N=6 leading to 15 possible PBAD's.
Oxygen displays for the principal contribution 102 (angle number 1, in red) a broad
distribution centred around the angle $\theta=135^o$, corresponding to the one defined
by the two closest (Ge) neighbours of oxygen, in agreement with experiments \cite{angle_geo2}.
Secondary distributions 103 and 203 show peaks centred at around $\theta\simeq$ 90$^o$ and 75$^o$.
In contrast, GeSe$_2$ exhibits a much sharper distribution for the same  102 contribution,
implying reduced  angular excursions as compared to GeO$_2$.
The contributions to the bimodal GeSe$_2$ distribution at 80$^o$ and
100$^o$ can be respectively assigned to edge-sharing and corner-sharing tetrahedra \cite{Masso_gese2},
a feature that is absent in oxides.
Six germanium centred angles (bottom panel) are found to have almost the same
distribution in GeO$_2$, centred at an angle of 109$^o$ typical of the tetrahedral environment.
\par
\begin{figure}
\begin{center}
\epsfig{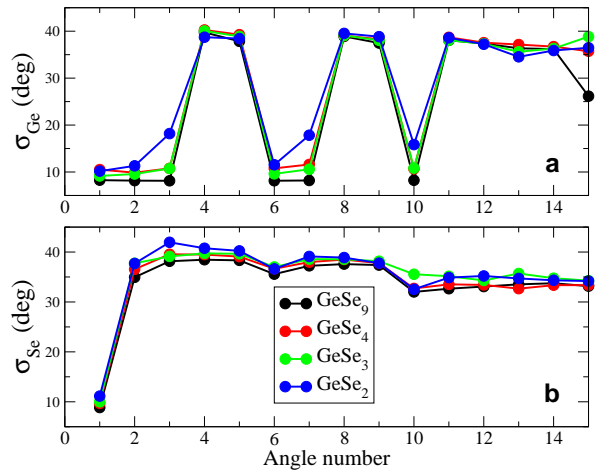}
\end{center}
\caption{(color online) Standard deviation $\sigma_{Ge}$ and $\sigma_{Se}$ extracted
from the partial bond angle distributions (PBAD) for four selected compositions in glassy Ge$_x$Se$_{1-x}$}
\end{figure}
The behavior of the standard deviations for the PBADs is shown in Fig. 2, where, for sake
of simplicity,  each PBAD is given a distinct number (see bottom part of Fig. 1 for the
definition).
For all chalcogenide or oxide systems, the PBADs relative to the Group IV (Si, Ge) atom
have a low standard deviation $\sigma_{\theta_{ij}}$, of the order of 10-20$^o$, for instance  
$\sigma_{\theta_{ij}}\simeq 7^o$ for the 102 PBAD of GeO$_2$.
These values are much smaller than those of the other distributions (105, 106, etc.),
found close at $\simeq$40$^o$.
In addition to very low angular excursions around
the tetrahedral angle of 109$^o$ (Fig. 2 top), oxides feature
all $\sigma_{\theta_{ij}}$ (i,j$<$4) nearly equal for the six relevant (Ge,Si) distributions.
A different situation occurs in stoichiometric chalcogenides (Fig. 2 bottom, red curve) 
which exhibit increased bending for the angles defining the tetrahedra (angle 3: 104, angle 7: 204, etc.).
These results exemplifies the difference in the bending nature of the tetrahedra in
these two families of networks, pointing to a higher rigidity of the tetrahedra in oxides, 
as all anglular excursions are maintained at the same low value (typically 7$^o$). 
\par
\begin{figure}
\begin{center}
\epsfig{figure=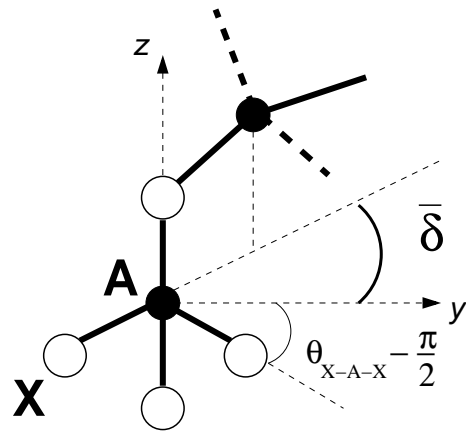,width=0.7\linewidth}
\end{center}
\caption{Definition of the dihedral angle $\bar \delta$ used in the text. Note that the upper black-filled atom can X (r$_X$=2) or A (r$_A$=4).}
\end{figure} 
Angle distributions around the Group VI atoms are markedly different.
Indeed, the selenium standard deviation $\sigma_{\theta_{12}}$ (angle number 1) is found to be low
(11.7$^o$ for GeSe$_2$) compared to the corresponding oxide system (26.7$^o$ for Ge$O_2$).
Therefore, the restoring effect associated with bending is much lower in
GeO$_2$, allowing for the identification of a {\it {broken}} angular constraint. One should also have in mind that the standard deviations 
found around 20$^o$ are close to those found in high temperature liquids (18-25$^o$ at 1373 K for GeSe$_2$), where angular constraints 
are assumed to be broken \cite{Mauro}.
This reduces the number density of constraints to n$_c$=0.33(${\frac {5}{2}}$r$_A$-3)+0.67(${\frac {5}{2}}$r$_X$-4)=3.00,
i.e. oxides are optimally constrained.
Because of its large fraction of ES tetrahedra \cite{sise2},
the situation of SiSe$_2$ appears to be somewhat intermediate
with $\sigma_{\theta_{12}}$=19.6$^o$, resulting from contributions
arising from both ES and CS tetrahedra which have respectively ineffective 
and intact Se bond-bending constraints. 
\par
\par
\begin{table}
\begin{center}
\begin{tabular}{cccccccc}
\hline\hline
System&SiO$_2$&GeO$_2$&SiSe$_2$&GeSe$_9$&GeSe$_4$&GeSe$_3$&GeSe$_2$\\ \hline
 & & & & & & & \\
$\bar \delta$(deg)&30.9&30.4&30.9&35.0&40.3&31.9&30.6\\
$\sigma_\delta$(deg)&17.7&17.5&23.9&27.4&25.9&20.6&21.6 \\ 
 & & & & & & & \\ \hline
\end{tabular}
\end{center}
\caption{Mean dihedral angle $\bar \delta$ and standard deviation $\sigma_\delta$ of
the dihedral angle distribution for the seven investigated systems}.
\end{table}
It is of interest to apply the above rationale to the Ge$_x$Se$_{1-x}$ family of systems,
since both the elastic nature (flexible, rigid) and the connectivity are strongly dependent on composition.
By increasing the Ge concentration the angular excursion
inside the tetrahedra is seen to increase (e.g. $\sigma_{Ge}$ moving up 
to 20$^o$ from less than 10$^o$ in the n=3 (104) PBAD) while leaving the stiffest angle (102) constant.
On the other hand, bending around the Se atoms is nearly unchanged and it does not display any noticable change
when the system becomes rigid.
To find a structural parameter pertaining to the Se atoms and sensitive to changes in 
composition, we have to resort to the dihedral angle $\bar \delta$ around a Se atom (see definition in Fig. 4).
As shown in Table I, the dihedral angular excursion takes a value of $\sigma_\delta$=27.4$^o$ 
for the flexible GeSe$_9$ composition and decreases significantly (down to 20$^o$-21$^o$) for 
the compositions of rigid systems GeSe$_3$ and GeSe$_2$.
Therefore, the network adapts to the predominant presence of Se atoms both by decreasing
the angular variability  inside the GeSe$_{4/2}$ tetrahedra, and by allowing for enhanced twisting  along
the Se chains. Finally, in view of the behaviour of $\sigma_A$ (A=O,Se) 
from both Figs. 2 and 3, it becomes clear that the angular motion inside SiO$_2$ and GeO$_2$ tetrahedra behaves very similarly 
to flexible GeSe$_9$ or optimally constrained GeSe$_4$ and GeSe$_3$ networks. 
On the other hand, these oxide systems definitely contrast with the corresponding stressed rigid chalcogenides SiSe$_2$ and GeSe$_2$.
\par
We showed that structural information gathered from molecular dynamics is able to provide
an atomic-scale counterpart to phenomenological constraint counting concepts applied to network glasses.
This leads to a clear picture of the topological differences between systems
having the same composition but different chemical nature and systems
made of the same species but differing in composition.
The differences between glasses (oxides and chalcogenides) of same stoichiometry
are rationalized in terms of amplitude  of the inter- and intratetrahedral bending angular variations.   
In oxides, the results show increased bending aroud the oxygen atom, consistent
with a direct Maxwell constraint counting and the non-stressed rigid nature of these glasses.
In systems undergoing a rigidity transition (e.g. Ge$_xSe_{1-x}$), it is found that
the angles defining the GeSe$_{4/2}$ tetrahedron soften with decreasing Ge weight 
while the Se angular bending is almost unchanged. Flexibility along the Se chains is
best accounted for in terms of twisting along the Se chains, found to increase with Se content.
The present approach finally provides a general framework for establishng bonding constraints in a neat way via model simulations, 
and should be used in the future for establishing constraint counting algorithms in more complex glassy materials.
\par
It is a pleasure to acknowledge ongoing discussions with
Christophe Bichara, Punit Boolchand and Jean-Yves Raty. This work has been supported by Agence Nationale de la Recherche (ANR) n.09-BLAN-0190-01.
The calculations were performed at the IDRIS computer center of CNRS and at the Computer Center DSI of UPMC (France).


\begin{thebibliography}{100}
       
\bibitem{Boolchand} {\em Insulating and Semiconducting glasses}, P. Boolchand Ed., World Scientific (Singapore, 2001).
\bibitem{Mysen_Richet} P. Richet and B. O. Mysen, {\em Silicate Glasses and Melts: Properties
and Structure} (Elsevier, Amsterdam, 2005).

\bibitem{Salmon2009} M. Wilson, P.S. Salmon, Phys. Rev. Lett. {\bf 103}, 157801 (2009).

\bibitem{Salmon1994} P.S. Salmon, Proc. R. Soc. A {\bf 445}, 351 (1994).

\bibitem{rigidity_book} {\em Rigidity theory and applications}, M.F. Thorpe and P.M Duxbury Eds.  (Kluwer Academic, Plenum Publishers New York, 1999).

\bibitem{Phillips1979} J.C. Phillips, J. Non-Cryst. Solids {\bf 34}, 153 (1979) ; M.F. Thorpe, J. Non-Cryst. Solids {\bf 57}, 355 (1983).

\bibitem{easily} M.C. Weinberg, D.R. Uhlmann and E.D. Zanotto, J. Am. Ceram. Soc. {\bf 72}, 2054 (1989).

\bibitem{Bool2001} P. Boolchand, X. Feng and W.J. Bresser, J. Non-Cryst. Solids {\bf 293}, 248 (2001).

\bibitem{Dove} K. Trachenko, M.T. Dove, V. Brazhkin, and F.S. El'kin, Phys.
Rev. Lett. {\bf 93}, 135502 (2004).

\bibitem{Wyart} M. Wyart, Ann. Phys. {\bf 30}, 1 (2005).

\bibitem{Science} M. Zhang and P. Boolchand, Science {\bf 266}, 1355 (1994).

\bibitem{PRE2006} M. Micoulaut, Y. Guissani and B. Guillot, Phys. Rev. E {\bf 73} 031504 (2006)

\bibitem{JCP1996} B. Guillot and Y. Guissani, J. Chem. Phys. {\bf 104}, 7633 (1996).

\bibitem{gese2} C. Massobrio, M. Micoulaut, P.S. Salmon, Solid State Sci. {\bf 12}, 199 (2010)

\bibitem{sise2} M. Celino and C. Massobrio, Phys. Rev. Lett. {\bf 90}, 125502 (2003).

\bibitem{Drabold} D. Tafen and D. A. Drabold, Phys. Rev. B{\bf  71}, 054206 (2005).

\bibitem{Gladden} R.W. Johnson, J. Non-Cryst. Solids {\bf 88}, 386 (1986).

\bibitem{Salmon} P.S. Salmon, A.C. Barnes, R.A. Martin and G.J. Cuello, J. Phys. Cond. Matt. {\bf 19}, 415110 (2007).

\bibitem{angle_geo2} R. Hussein, R. Dupree and D. Holland, J. Non-Cryst. Solids {\bf 246}, 159 (1999).

\bibitem{Masso_gese2} C. Massobrio and A. Pasquarello, Phys. Rev. B{\bf 77}, 144207 (2008).

\bibitem{Mauro} P.K. Gupta and J.C. Mauro, J. Chem. Phys. {\bf 130}, 094503 (2009).


\end{thebibliography}
\end{document}